\documentclass[prb,twocolumn,amsmath2000,superscriptaddress]{revtex4}
\usepackage[dvips]{graphicx} 
\usepackage{psfrag}
\usepackage{wasysym}
\usepackage{color}
\definecolor{Blue}{rgb}{0.00, 0.00, 1.00}
\definecolor{Red}{rgb}{1.00, 0.00, 0.00}
\begin{document}

\newcommand{\fig}[2]{\includegraphics[width=#1]{#2}}

 
\title{Fractionalization in an Easy-axis Kagome Antiferromagnet}


\author{L. Balents}
\affiliation{Department of  Physics, University of California,
Santa Barbara, CA 93106--4030}
\author{M. P. A. Fisher}
\affiliation{Institute for Theoretical Physics, 
University of California, Santa Barbara, CA 93106--4030}
\author{S. M. Girvin}
\affiliation{Institute for Theoretical Physics, 
University of California, Santa Barbara, CA 93106--4030}
\affiliation{Yale University, Physics Department, Sloane Physics Lab, New
Haven, CT 06520-8120} 

\date{\today}

\begin{abstract}
  We study an antiferromagnetic spin-$1/2$ model with up to third
  nearest-neighbor couplings on the Kagome lattice in the easy-axis
  limit, and show that its low-energy dynamics are governed by a four
  site XY ring exchange Hamiltonian.  Simple ``vortex pairing''
  arguments suggest that the model sustains a novel fractionalized
  phase, which we confirm by exactly solving a modification of the
  Hamiltonian including a further four-site interaction.  In this
  limit, the system is a featureless ``spin liquid'', with gaps to all
  excitations, in particular: deconfined $S^z=1/2$ bosonic ``spinons''
  and Ising vortices or ``visons''.  We use an Ising duality
  transformation to express vison correlators as non-local strings in
  terms of the spin operators, and calculate the string correlators
  using the ground state wavefunction of the modified Hamiltonian.
  Remarkably, this wavefunction is exactly given by a kind of
  Gutzwiller projection of an XY ferromagnet.  Finally, we show that
  the deconfined spin liquid state persists over a finite range as the
  additional four-spin interaction is reduced, and study the effect of 
  this reduction on the dynamics of spinons and visons.
\end{abstract}

\maketitle

\vspace{0.15cm}



\section{Introduction} 
It has now been almost 15 years since P.W. Anderson suggested that
two-dimensional (2d) spin one-half antiferromagnets might condense
into a featureless ``spin liquid'' quantum ground state\cite{PWA}.  In
close analogy with the one-dimensional Heisenberg antiferromagnetic
chain, the 2d spin liquid was posited to support deconfined spinon
excitations -- ``particles'' carrying s=1/2 in stark contrast with the
s=1 triplet excitations of more familiar non-magnetic phases such as
the spin-Peierls state and with the $S^z=1$ magnon excitations of the
2d Ne\'el state\cite{Sachdevbook}.  Early attempts to demonstrate the
existence of the 2d spin liquid focussed on quantum dimer
models\cite{RKdimer} motivated directly by Resonating Valence Bond
(RVB) ideas\cite{RVB}, slave-Fermion mean field theories\cite{slave}
and large N generalizations\cite{LargeN} of the spin models.  While
the topological character of the spin liquid was mentioned in some of
these pioneering studies\cite{oldtop}, generally the focus was on
characterizing the spin liquid by an absence of spin ordering and
spatial symmetry breaking.  In the past few years, it has been
emphasized that the precise way to characterize a 2d spin liquid
phase\cite{newtop} -- as with other 2d fractionalized phases -- is in
terms of ``topological order,'' a notion introduced by X.-G. Wen in the
context of the fractional quantum Hall effect\cite{Wentop}.  Central
to the notion of topological order in 2d is the presence of
vortex-like excitations with long-ranged statistical
interactions\cite{oldtop,Z2}.  In the simplest 2d spin liquid these
point-like excitations have been dubbed ``visons'' since they carry an
Ising or $Z_2$ flux\cite{Z2}.  Upon transporting a spinon around a
vison, the spinon's wavefunction acquires a minus sign.  A theoretical
description of this long-ranged statistical interaction is most
readily incorporated in the context of a gauge theory with a discrete
Ising symmetry, in which the visons carry the $Z_2$ flux and the
spinons the $Z_2$ charge\cite{Z2}.  The $Z_2$ gauge theory can be
dualized into a vortex representation, wherein the topological order
follows from the notion of ``vortex pairing''\cite{vortpair}.

Efforts to identify microscopic spin Hamiltonians that might actually
exhibit such topologically ordered phases have focussed on strongly
frustrated 2d s=1/2 antiferromagnets.  Due to the ``sign problem''
these efforts have been essentially limited to exact diagonalization
studies on very small lattices.  Nevertheless, such numerics do
identify a few models which appear to be in a spin liquid phase: the
Kagome antiferromagnet with near neighbor interactions\cite{kagnum}
and a triangular lattice model with 4-spin ring exchange
terms\cite{ringnum}.  The importance of multi-spin ring exchange
processes in driving 2d fractionalization is also apparent within the
$Z_2$ gauge theory formulation\cite{Z2}.  In an important recent
development, Moessner and Sondhi\cite{sondhi} have compellingly argued
that a particular quantum dimer model on the triangular lattice is in
a featureless liquid phase, closely analogous to a the desired ``spin
liquid" phase of a spin Hamiltonian.

In this paper we re-visit the s=1/2 Kagome antiferromagnet, in the
presence of second and third neighbor exhange interactions.  By
passing to an easy-axis limit of this model, substantial analytic and
numerical progress is possible both in establishing the presence of a
fractionalized spin liquid and of directly analyzing it's topological
properties.  Specifically, in the easy-axis limit we map the model
exactly onto an XY Hamiltonian consisting solely of a local 4-spin
ring exhange interaction.  Since the sign of the ring exchange term is
``bosonic" -- opposite to the sign obtained upon cyclically permuting
four underlying s=1/2 fermions (eg. electrons)\cite{ringnum} -- 
the Hamiltonian does
{\it not} suffer from a sign problem and so should be amenable to
quantum Monte Carlo.  Furthermore, if the two levels of the spin-$1/2$
on each site of the Kagome lattice is reinterpreted as the
presence or absence of a (quantum) dimer living on a
bond of a triangular lattice, the model can be reinterpreted as a
quantum dimer model which is very similar to that considered by
Moessner and Sondhi\cite{sondhi}, the distinction being that three, rather than
one, dimers emerge from each site.  This realization allows us to
exploit the important work of Rokhsar and Kivelson\cite{RKdimer} who identified an
exactly soluble point of a generalized square lattice quantum dimer
model.  With a similar generalization, our model also possesses an
exact zero energy wavefunction: an equal weight superposition of all
allowed spin configurations in the low energy singlet sector.  We show
that this wavefunction can be viewed as an exact version of the
popular variational state consisting of the Gutzwiller projection of a
superfluid/superconductor\cite{Gutzwiller}.  Finally, 
we are able to implement an exact
duality transformation which enables us to identify the operators
which create both the spinon excitation and the topological vison
excitation.  Employing the exact wave function, we compute numerically
the vison 2-point correlation function, and show that it is
exponentially decaying - the hallmark of a 2d fractionalized phase\cite{Z2}.
We thereby demonstrate that the (gapped) spinons are genuine
deconfined particle-like excitations.

The paper is organized as follows.  In Section II we introduce a generalized
$s=1/2$ Kagome antiferromagnet and show how it can be mapped onto a
bosonic ring model in the easy-axis limit.
With a slight further generalization, we identify an
exactly soluble point in Section III and obtain
an exact spin-liquid ground state wavefunction.
In Section IV we exploit an exact duality transformation
which maps the Kagome spin model onto a $Z_2$ gauge theory
living on the dual lattice to identify the spinon and vison excitations.
The vison two-point correlation function is then evaluated numerically
using the exact wavefunction in Section V, and we demonstrate
that it is short-ranged thereby directly establishing
the presence of fractionalization in the spin-liquid ground state.
Finally, Section VI is devoted to a brief discussion
of the implications of this finding.

\section{Model}

\begin{figure}[hbt]
\psfrag{J1}{$J_1$}
\psfrag{J2}{$J_2$}
\psfrag{J3}{$J_3$}
\psfrag{S1}{1}
\psfrag{S2}{2}
\psfrag{S3}{3}
\psfrag{S4}{4}
\psfrag{A1}{$\vec{a}_1$}
\psfrag{A2}{$\vec{a}_2$}
\centerline{\fig{8cm}{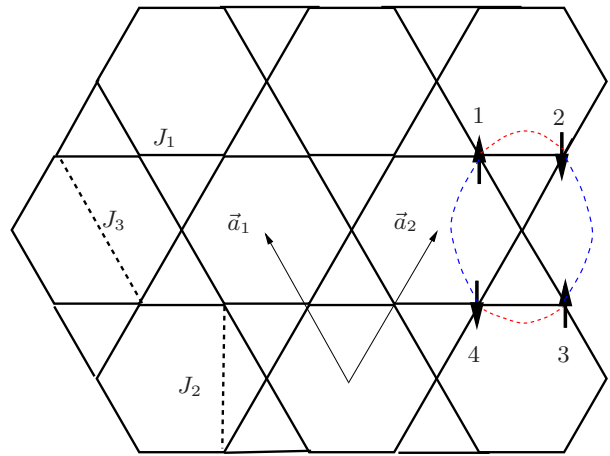}}
\caption{Kagome lattice and interactions.  Two primitive vectors
  $\vec{a}_1,\vec{a}_2$ are shown, as are the labels $1\ldots 4$ for
  the four sites on a bow-tie.  The ring term is generated both from
  the red (short dashes) and blue (long dashes) virtual exchange processes.}   
\label{fig:lattice}
\end{figure}

We consider a spin one-half Heisenberg antiferromagnet on a Kagome
lattice with Hamiltonian
\begin{equation}
{\cal H} = \sum_{ij} J_{ij} \vec{S}_i \cdot \vec{S}_j   .
\end{equation}
Since the Kagome lattice consists of corner sharing triangles, the
nearest neighbor exchange interaction, denoted $J_1$, is strongly
frustrating.  Here we extend this standard near-neighbor model to
incude further neighbor interactions, $J_2, J_3$, which act between
pairs of sites on the hexagons in the Kagome lattice (Fig.~\ref{fig:lattice}).
Specifically, two spins on the same hexagon separated by 120 degrees
are coupled via $J_2$, and $J_3$ is the coupling between two spins
diametrically across from one another on the hexagon.

Instead of the usual nearest-neighbor Kagome Antiferromagnet (with $J_2=J_3=0$), we specialize instead to the case with equal
exchange interactions, $J_1=J_2=J_3=J$.  This generalized Kagome
Antiferromagnet can be cast into a simple form:
\begin{equation}
  {\cal H} = J \sum_{\mbox{\small\hexagon}} \vec{S}_{\mbox{\small\hexagon}} \cdot \vec{S}_{\mbox{\small\hexagon}},
  \label{eq:hex}
\end{equation}
where the summation is over all hexagons on the Kagome lattice and
$\vec{S}_{\mbox{\small\hexagon}} = \sum_{i=1}^6 \vec{S}_i$ is the sum of the
six spins on each hexagon.  A similar form was obtained by Palmer and
Chalker\cite{Chalker} for a Heisenberg model on the ``checkerboard''
lattice, with the Hamiltonian expressed as a sum over the total spin
living on elementary square plaquettes, squared.

As for the nearest-neighbor model, the generalized Kagome antiferromagnet described by Eq.~\ref{eq:hex}\ has a
non-trivial classical limit.  There is a thermodynamically large set
of classical ground states, which includes any configuration for which
the classical vector $\vec{S}_{\mbox{\small\hexagon}}=0$ for each hexagon.  The
breaking of this degeneracy by quantum fluctuations could give rise to
``order-by-disorder''.  For the spin-$1/2$ case of interest, however,
Eq.~\ref{eq:hex}\ is essentially intractable analytically.  
To make progress, we retain SU(2) spins with $S=1/2$ on each site, but
generalize the Hamiltonian to allow for anisotropic exchange
interactions.  Specifically, we consider an ``easy axis" limit, with
the exchange interaction along the $z-$axis in spin space larger than
in the $x-y$ plane: $J_z > J_\perp$.  In the extreme easy-axis limit,
one can first analyze the $J_z$ terms alone, and then treat the
remaining terms as a perturbation: ${\cal H} = {\cal H}_0 + {\cal H}_1$
with,
\begin{equation}
{\cal H}_0 = J_z \sum_{\mbox{\small\hexagon}} \left( S^z_{\mbox{\small\hexagon}} \right)^2   ,
\end{equation}
and
\begin{equation}
{\cal H}_1 = J_\perp \sum_{\mbox{\small\hexagon}} \left[\left( S^x_{\mbox{\small\hexagon}} \right)^2 +
\left( S^y_{\mbox{\small\hexagon}} \right)^2 -3\right] ,
\end{equation}
where the subtraction of $3$ was included for convenience.  In an
eigenbasis of $S^z_i=S=\pm 1/2$, the Hamiltonian ${\cal H}_0$
describes a classical spin system.  The classical ground state
consists of all spin configurations which have zero ($z$-axis)
magnetization on each and every hexagonal plaquette: $S^z_{\mbox{\small\hexagon}}
=0$.  There are many such configurations (note that unlike the nearest-neighbor model,
the generalized Kagome antiferromagnet is unfrustrated in the easy-axis limit), with a ground state
degeneracy that grows exponentially with system size, much like other
fully frustrated classical spin models such as the triangular lattice
Ising antiferromagnet.  The full Hamiltonian, ${\cal H}$, lifts this huge
degeneracy, splitting the classically degenerate ground states into a
{\sl low-energy manifold}, still characterized, however, by the good
quantum numbers $S^z_{\mbox{\small\hexagon}}=0$.  

Some properties of this easy-axis limit are immediately evident.  
For instance, all states in the low-energy manifold
have $S^z_{\mbox{\small\hexagon}}=0$ for every hexagon, and there is a large gap of
approximately $J_z$ to states with any non-zero $S^z_{\mbox{\small\hexagon}}$.
Hence the ground state has in this sense a ``spin gap''.  Thus the
easy-axis generalized Kagome antiferromagnet has no XY
spin order, but translational symmetry breaking is not precluded.
More subtle aspects of this model are less evident.
In particular, we would like to ascertain the presence or
absence of more subtle ``topological'' order, and the types of
``singlet'' (more precisely $S^z=0$) and spinful ($S^z \neq 0$)
excitations.

To proceed, we treat ${\cal H}_1$ as a perturbation with $J_\perp \ll
J_z$, and project back into this {\sl low energy manifold} of
degenerate classical ground states with $S^z_{\mbox{\small\hexagon}} =0$.  (This
procedure is very much analogous to the derivation of the Heisenberg
model starting from the Hubbard model with $t\ll U$.  Indeed, in the
language of ``hard core bosons" in which the boson number corresponds
to $S^z_i +1/2$, the perturbing Hamiltonian ${\cal H}_1$ describes
boson hopping amongst a pair of sites on the same hexagon.)  Within
second order degenerate perturbation theory (in $J_\perp$) for the
low-energy manifold, there are two types of (virtual) processes which
contribute, preserving the vanishing magnetization on every hexagon.
In the first, two antiparallel spins within a single hexagon exchange
and then exchange back again.  This ``diagonal" process leads (within
the low-energy manifold) to a simple constant energy shift
$E_0=-(9/2)N_{\mbox{\small\hexagon}} J_\perp^2/J_z $, where $N_{\mbox{\small\hexagon}}$ is the
total number of hexagons.  Because this trivial shift does not split
the extensive degeneracy, we neglect it in what follows.  More
interesting are off-diagonal processes, in which two pairs of
antiparallel spins on opposite sites of a 5-site ``bow-tie" plaquette
exchange (see Fig.~\ref{fig:lattice}).  This process involves
spins on only four
sites, and is an analog of electron exchange ``ring'' moves.  One can
readily verify that such ``ring'' moves on the bow-tie leave invariant
the ($z$-axis) magnetization on every hexagon.

Up to second order in $J_\perp/J_z$, within the low-energy manifold,
the full Kagome Heisenberg antiferromagnet is thereby reduced to the
form: ${\cal H}_0 + {\cal H}_{ring}$ with
\begin{equation}
  {\cal H}_{\rm ring} = -J_{\rm ring} \sum_{\bowtie} ( S^+_1 S^-_2
  S^+_3 S^-_4 +  {\rm h.c.} )  , \label{eq:Sring}
\end{equation}
where the labels $1\ldots 4$ denote the four spins at the ends of each 
bowtie as labeled in Fig.~\ref{fig:lattice}.  
Here the ring exchange interaction $J_{\rm ring} = J_\perp^2 /J_z$, and by
assumption one has $J_\perp \ll J_z$.  It is noteworthy that in this
extreme easy-axis limit the frustrated Kagome magnet does {\it not}
have a sign problem, and as such could be profitably attacked via
quantum Monte Carlo.  

\section{Soluble Spin Liquid}

We now use ${\cal H}_{\rm ring}$ to address the nature of the spin-gapped
state of the easy-axis generalized Kagome antiferromagnet.  Several arguments
point to a spin-liquid phase which supports fractionalized
``spinon'' excitations which carry spin $S^z=1/2$.  Such a
fractionalized state must also support vortex-like excitations, dubbed
``visons'', which carry no spin but have a long-ranged
statistical interaction with spinons.  

A first suggestion to this effect comes from viewing ${\cal H}_{\rm
ring}$ as a lattice boson model, and a spin liquid state thereby as
a bosonic Mott insulator.  Generally, such
bosonic insulating states can be regarded as quantum-mechanical condensates
of vortices\cite{MPAFvortcond}.  To examine the vortex excitations, it is convenient
to think of $S_i^\pm$ as lattice boson
raising and lowering operators.  Formally, one may then express
$S_i^\pm = e^{\pm i \phi_i}$ -- fluctuations in the $U(1)$ phases
$\phi_i$ (conjugate to $S_i^z$) are induced by the constraint
$S_i^z=\pm 1/2$.  It is then illuminating to re-express the bosonic
ring term as
\begin{equation}
  {\cal H}_{\rm ring} = - 2 J_{\rm ring}\sum_{\bowtie}
  \cos(\phi_1 - \phi_2 +\phi_3 - \phi_4) . \label{eq:pring}
\end{equation}
Consider now a vortex centered on some site (the ``core'').
Classically, for the four sites on the bow-tie surrounding the vortex
core, $\phi_j = (j/4) 2\pi N_v $, where $N_v$ denotes the number of
vortices (vorticity) on this plaquette.  The (core) energy of this
vortex configuration is proportional to
\begin{equation}
  E_{\rm vort} = 2J_{\rm ring} ( 1 -  \cos(N_v \pi)) .
\end{equation}
Notice that plaquettes with an odd number of vortices, $N_v$, cost an
energy $4J_{\rm ring}$ relative to the even-$N_v$ plaquettes.  In
particular, a single strength vortex is costly, but double-strength
vortices are cheap.  The same conclusion can be shown more formally
using an exact duality transformation.  

Typically single strength vortices condense, but one can also imagine
insulating states which result from a condensation of composites made
from $N_v$ vortices\cite{vortpair}.  Such insulators are necessarily
fractionalized since they support deconfined (but gapped) charge
excitations with ``boson charge'' $Q=S^z=1/N_v$.  Based on the
energetics of the ring term which tends to expel single vortices with
double vortices being energetically cheaper, one expects that the
insulating state for the Kagome ring model will have spin $S^z=1/2$
excitations -- if it is fractionalized at all.  If fractionalized, the
``vison'' can be understood as an unpaired vortex state in the
vortex-pair condensate, a ``dual'' analog of a BCS quasiparticle.

Further evidence that the ground state of this model might be
fractionalized comes from its formal equivalence to a particular
quantum dimer model.  Mapping to a dimer model is straightforward
since the sites of the Kagome lattice can be viewed as the centers of
the links of a triangular lattice.  The two $S^z = +(-) 1/2$ states on
a site correspond to the presence (or absence) of a dimer on the
associated link on the triangular lattice.  The ring term above
corresponds directly to the elementary quantum-dimer move on the
triangular lattice considered recently by Sondhi and Moessner\cite{sondhi}.  The
only difference with the standard dimer model is that in this instance
there are {\sl three} dimers coming out of every site of the
triangular lattice instead of the usual one.  Sondhi and Moessner
considered an additional ``diagonal" term (see below) in the triangular
lattice quantum dimer model, and argued that the model was in a spin
liquid state in portions of the phase diagram.  Central to their
argument was an exactly soluble point of the model, first exploited by
Rokhsar and Kivelson (RK)\cite{RKdimer} in the square lattice quantum dimer model.  The
additional term is diagonal in $S^z_i$, and may be written ${\cal
  H}_{\rm nf} = u_4 \sum_{\rm r \in \bowtie} \hat{P}_{\rm flip}(r)$, where
\begin{equation}
  \hat{P}_{\rm flip}(r) = \sum_{\sigma=\pm1}\prod_{j\in r =1}^4
  ({\frac{1}{2}} +\sigma (-1)^j S^z_j) .
\end{equation}

The operator $\hat{P}_{\rm flip}(r)$ is a projection operator onto the two
flippable states of the bow-tie ring $r$.  This term in the
Hamiltonian can be combined with ${\cal H}_{ring}$ and written in the
suggestive form:
\begin{equation}
  {\cal H}_{\rm ring} + {\cal H}_{\rm nf} =  \sum_{r}
  \hat{P}_{\rm flip}(r)\lbrace -J_{\rm ring} \prod_{j=1}^4  2 S^x_j  +
  u_4 \rbrace. 
\label{HamRK}
\end{equation}

When $u_4=J_{\rm ring}$ one can write down exact ground state(s) which
have the product of $2S_x$ equalling one on all bow-tie rings.  One
such state is the $XY$ ferromagnet with $S_j^x=1/2$ on every site.  In
the hard-core boson description, this corresponds to a superfluid
state (albeit an unusual one with no zero-point fluctuations).  One
must project back into the subspace in which there are three bosons on
every hexagon ($S^z_{\mbox{\small\hexagon}}=0$), since otherwise this state
will not be an eigenstate of ${\cal H}_0$.  (Actually, several
distinct projections are generally possible, onto different sectors
disconnected from one another under the action of ${\cal H}_{\rm
  ring}$.  These give degenerate ground states.) This projection of a
superfluid wavefunction to obtain a bosonic insulating state is
analogous to the Gutzwiller projections of superconducting
wavefunctions to obtain variational states for quantum spin
models\cite{Gutzwiller}, but there is an important difference.  In the
present instance, the constraints (of 3 bosons on every hexagon) {\it
  commute} with the Hamiltonian ${\cal H}_{ring}$ which hops the
bosons, in contrast to the no-double occupancy constraint which does
not commute with the electron kinetic energy term in Hubbard type
models.  Thus, in our case the wave function after projection is still
an exact eigenstate of the full Hamiltonian.

\section{Duality, Visons, and Spinons}

\begin{figure}[hbt]
\centerline{\fig{8cm}{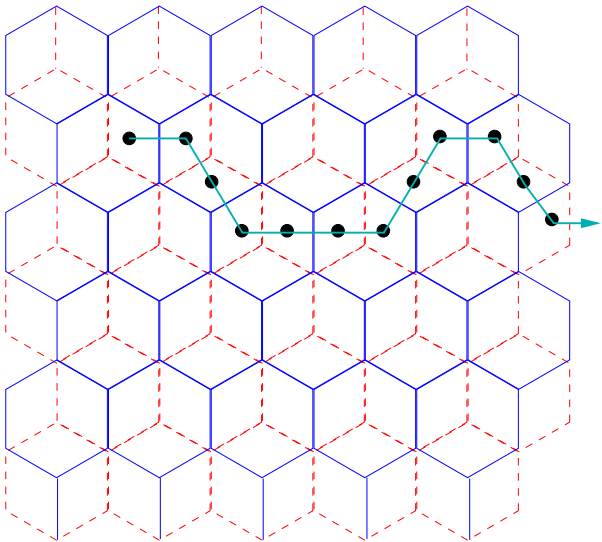}}
\caption{Dice lattice shown as two interpenetrating honeycombs,
  indicated by blue (solid) and red (dashed) lines.  A blue vison is
  created geometrically by multiplying $2S^z_i$ over the underlying
  kagome sites (centers of parallelograms, shown by solid dots)
  through which the ``string'' indicated passes.  In the dual
  variables, this product is given by the product of blue gauge fields
  $\sigma_{ij}^x$ cut by the string shown.  The ``blueness'' of the
  vison shown owes to the fact that only a single spin $S_i^z$ is
  contained within the originating blue hexagon.}
\label{fig:dice}
\end{figure}

Before studying this wave function, it is convenient to expose the
vison degrees of freedom via a duality transformation.  Specifically,
we will employ the standard $2+1$ dimensional Ising duality which
connects a global spin model to a $Z_2$ gauge theory with gauge fields
living on the links of the dual lattice\cite{Kogut}.  In our case the
global spin model is the Kagome model ${\cal H}_{ring} + {\cal
  H}_{nf}$ in Eq.~\ref{HamRK}, so that the dual lattice is the ``dice"
lattice, which can conveniently be constructed in terms of two
interpenetrating honeycomb lattices as depicted in
Fig.~\ref{fig:dice}.  On the operator level, the duality
transformation is implemented by re-expressing $S^x$ and $S^z$
directly in terms of the dual gauge fields, $\sigma_{ij}^\mu$ - a set
of Pauli matrices living on the links of the dice lattice:
\begin{equation}
S_i^x = {\frac{1} {2}} \prod_{jl \in \diamond} \sigma_{jl}^z ,
\label{eq:sxdual} 
\end{equation}
and
\begin{equation}
S_i^z = {\frac{1}{2}}
\prod_{jl=i}^{\infty}\!\!\!\!\!\!\!\!\!\!\longrightarrow \sigma_{jl}^x 
\label{dualityz} . 
\end{equation}

Here, the first product is taken around an elementary 
four-sided plaquette on the dice lattice
which surrounds the spin $S_i^x$.  The second product involves
an infinte string which connects sites of the Kagome lattice,
eminating from the site $S_i^z$ and running off to spatial infinity.
For every bond of the dual dice lattice which is bisected by this string,
a factor of $\sigma_{ij}^x$ is present in the product.
To assure that this definition is independent of the
precise path taken by the string, requires
imposing the constraint that the product of $\sigma_{ij}^x$
on all bonds connected to each site on the dice lattice
is set equal to unity:
\begin{equation}
{\cal G}_i = \prod_{\langle ji\rangle} \sigma_{ij}^x = 1  ,
\label{gaugeop}
\end{equation}
where here $j$ labels the near-neighbor sites to $i$.  These local
$Z_2$ gauge constraints must be imposed on the Hilbert space of the
dual theory.  In the resulting dual gauge theory, these constraints
are analogous to Coulomb's law ($\nabla \cdot E =0$) in conventional
electromagnetism.  The necessity of including the constraints can be
simply seen by counting degrees of freedom: there are twice as many
(six) bonds per unit cell on the dice lattice as sites (three) per
unit cell on the Kagome lattice, hence to maintain the physical
Hilbert space of the original spins (site variables) requires
restricting the gauge fields (bond variables).

The dual Hamiltonian takes the form:
\begin{equation}
{\cal H}_{dual}  = \sum_r \hat{P}_{\rm flip}(r) \lbrace -J_{\rm ring}
\prod_{{\color{Red} \hexagon_r}} \sigma^z 
\prod_{{\color{Blue} \hexagon_b}} \sigma^z + u_4 \rbrace  ,
\end{equation}
where the products are taken around the two hexagonal plaquettes
of the dice lattice which surround the given Kagome site $r$.
These products measure ``magnetic flux'' (in the dual
gauge fields) through hexagons belonging to
the two honeycomb sublattices.  The flip term becomes,
\begin{equation}
  \hat{P}_{\rm flip}(r) = \prod_{j \in r =1}^4 (1 - \sigma_{ij}^x
  \sigma_{jl}^x ), 
\end{equation}
where the product is taken over pairs of bonds on the elementary
dice plaquette which both connect to the same site, $j$.

One can readily verify that the operators which implement a local
gauge transformation, ${\cal G}_i$ in Eqn.~\ref{gaugeop}, commute with
this dual Hamiltonian.  Equivalently, since ${\cal G}_i \sigma_{ij}^z
{\cal G}_i = - \sigma^z_{ij}$, the dual Hamiltonian is invariant under
the general $Z_2$ gauge transformation,
\begin{equation}
\sigma^z_{ij} \rightarrow \epsilon_i \sigma^z_{ij} \epsilon_j  ,
\end{equation}
with arbitrary $\epsilon_i = \pm 1$.  Remarkably, though, it turns out
that this gauge theory actually has an additional set of local $Z_2$
symmetries.  In particular, it is possible to transform the $\sigma^z$
gauge fields living on the blue (or red) links separately, and still
leave the Hamiltonian invariant.  Eqivalently, one can define local
red or blue gauge operators which commute with the Hamiltonian:
\begin{equation}
{\cal G}^r_i = \prod_{{\color{Red}\langle ij\rangle_r}} \sigma^x_{ij} ,
\end{equation}
with the product over red links which eminate from site $i$, and
similarly for the blue links.  On the dice lattice, for each 6-fold
coordinate site there corresponds both a blue and red local gauge
operator, whereas the 3-fold coordinated sites are either red or
blue.

The presence of this additional local symmetry can be directly traced
to the conservation of the magnetization $S^z_{\mbox{\small\hexagon}}$ on each hexagon
of the original Kagome lattice ring spin-model (note that this is the
conservation of dimer number emerging from each site on the equivalent 
triangular lattice dimer model).  Indeed, upon using
Eqn.~\ref{dualityz}, one can show that for the 6-fold coordinate sites
of the dice lattice
\begin{equation}
{\cal G}_i^{red} = {\cal G}_i^{blue} = - \exp \left[i \pi
  S^z_{{\mbox{\small\hexagon}},i}\right] , 
\end{equation}
where the center of the hexagon is at site $i$ on the dice lattice.
The right hand side of this expression can be interpreted as a $Z_2$
``charge" living on the 6-fold coordinate sites of the dice lattice,
since it equals the (lattice) divergence of the $Z_2$ ``electric
fields".  For the ``singlet" sector of the theory with
$S^z_{\mbox{\small\hexagon}} = 0$ for all hexagons, the right hand side is
simply minus one.  But more generally, this expression indicates that
hexagons with a non-zero (odd integer) value of the globally conserved
spin, $S^z_{\mbox{\small\hexagon}}= \pm 1$, also carry both a red and a blue
$Z_2$ gauge charge.

This fact allows us to identify both the spinon and vison excitations
in the theory.  Specifically, consider starting in the ``singlet"
sector of the theory with zero magnetization on every hexagon of the
Kagome lattice, and flipping a single spin.  Since each site of the
Kagome lattice is shared by two hexagons, this creates two hexagons
each with $S^z_{\mbox{\small\hexagon}} =1$.  By adding a small near
neighbor spin exchange it is possible to hop these two magnetized
hexagons, and to spatially separate them.  As we demonstrated above,
such magnetized hexagons also carry both a red and a blue $Z_2$
charge.  Provided the dual gauge theory is in it's deconfined phase,
these magnetized hexagons can propagate as independent particles.
Since two such magnetized hexagons were created when we added spin one
to the system (by flipping the single spin), each magnetized hexagon
must carry spin $S^z = 1/2$, and we can thereby identify these
excitations as the deconfined spinons.

A 2d spin liquid with deconfined spinons must necessarily support
topological vortex like excitations -- the visons\cite{Z2}.  The vison acts as
a source of $Z_2$ ``flux'' for the spinon, whose wave function changes
sign as it is transported around a vison.  In the $Z_2$ gauge theory
formulation of 2d fractionalization, the flux of the vison corresponds
generally to a plaquette with $\prod_{\rm plaq} \sigma^z =-1$.  Since
the spinons which hop on the 6-fold coordinated sites of our dice
lattice carry both a red and a blue $Z_2$ gauge charge, it is clear
that this spin liquid phase will support two flavors of visons - a red
(blue) vison corresponding to a flux penetrating one red (blue)
hexagon of the dice lattice.

Due to the long-ranged statistical interaction between visons and
spinons, it is not possible to have both excitations present and
freely propagating.  In particular, if the visons are gapped
excitations they will be expelled from the ground state and the spinons
will be deconfined.  On the other hand, a proliferation and
condensation of visons will lead to spinon confinement.  Thus, in
order to establish 2d fractionalization it is adequate to show an
absence of vison condensation.  A useful diagnostic for this is the
vison 2-point correlation function:
\begin{equation}
V (r_i - r_j) = \langle \hat{v}_i \hat{v}_j \rangle   ,
\end{equation}
where $\hat{v}$ denotes a vison creation operator.  When this correlation
function is short-ranged, the visons are not condensed, and the system
is fractionalized.

In order to evaluate this correlation function for the Kagome spin
model, it is necessary to express this vison 2-point function in terms
of the original spin operators.  To this end, we first note that from
the definition in Eq.~\ref{dualityz}, it is apparent that the operator
$2S^z_i$ creates both a red and a blue vison, $2S^z_i = \hat{v}_i^{r}
\hat{v}_i^{b}$, since it introduces a $Z_2$ flux through the red and
blue hexagons of the dice lattice which enclose the spin.  Since
$\hat{v}_i \hat{v}_i =1$, a single (red) vison (say) can be created by
stringing together an infintely long product of spin operators $S^z$.
This ``string'' starts at the given site of the Kagome lattice and
joins neighboring spins making only $\pm 30^0$ turns eventually
running off to spatial infinity, but otherwise is arbitrary.
Explicitly, the vison two-point function is then
\begin{equation}
  V_{ij} = \left|\langle 0| 
    \prod_{k=i}^{j}\!\!\!\!\!\!\!\!\!\!\longrightarrow  2S_k^z
    |0\rangle\right|, \label{eq:sijdef}
\end{equation}
where $|0\rangle$ denotes the ground state, and the product in
Eq.~\ref{eq:sijdef}\ is taken, as described above, along some path on
the Kagome lattice starting at site $i$ and ending at site $j$,
containing an even number of sites, and making only ``$\pm 30^\circ$''
turns left or right.  Due to the constraint $S^z_{\rm hex}=0$, the
latter product is path-independent up to an overall sign (hence the
absolute value in Eq.~\ref{eq:sijdef}).  We also define for
convenience the physically interesting spin-spin correlator,
\begin{equation}
  C_{ij} = \langle 0| S_i^z S_j^z |0\rangle, \label{eq:cijdef} 
\end{equation}

\section{Correlators at the Soluble Point}
 
With $V$ and $C$ defined appropriately in terms of the spins, we are
now in a position to evaluate them using the exact RK wavefunction.
Specifically, we consider exact ground state
wavefunctions (at the RK point) on the torus defined by identifying
sites connected by the two winding vectors $\vec{W}_1= n_1 \vec{a}_1$
and $\vec{W}_2 = n_2 \vec{a}_2$, where $n_1,n_2$ are positive even
integers, and $\vec{a}_1,\vec{a}_2$ are primitive vectors (see
Fig.~\ref{fig:lattice}).  The degeneracies etc. of such wavefunctions
are nearly identical to that discussed by Moessner and Sondhi, so we
do not go into detail here.  We focus on the wavefunction $|0\rangle$,
which has been projected onto a single topologically connected sector.

\begin{figure}[hbt!]
\psfrag{String Correlator}{$V$}
\psfrag{Spin Correlator}{$C$}
\psfrag{|xi-xj|}{${\rm min}|\vec{x}_i-\vec{x}_j|$}
\psfrag{ln(Vij)}{$\ln V$}
\psfrag{ln(Cij)}{{\color{Red} $\ln C$}}
\centerline{\fig{9cm}{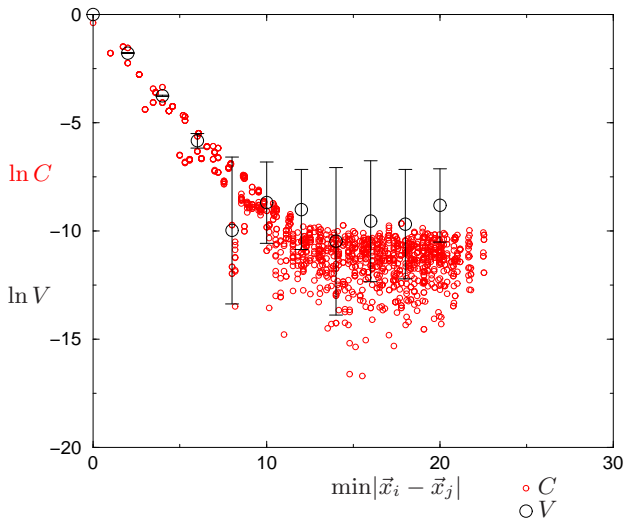}}
\caption{Correlators for the $(20,20)$ Kagome torus.  The horizontal axis
  ${\rm min}|\vec{x}_i-\vec{x}_j|$ is the shortest distance between
  sites $i$ and $j$ on the torus, in units of the inter-site distance.
  Large circles with error bars show the logarithm of $V_{ij}$,
  while small circles represent $\ln(C_{ij})$.  Each small 
  circle represents an average over all pairs of sites with a fixed
  separation, while the large circles with error bars represent the
  distribution of string correlators of pairs of sites connected by a
  {\sl horizontal} string.  Both correlators clearly decay
  exponentially, with apparently the same exponent, albeit with larger
  numerical errors for the string correlator, due presumably to the
  fact that this data is less spatially averaged.  }
\label{fig:numerical}
\end{figure}

The expectation values of interest can be evaluated stochastically
using a {\sl classical} ``infinte temperature'' Monte Carlo algorithm,
which ``random walks'' through the various components of the
wavefunction.  Our numerical results for a torus with $n_1=n_2=20$ are
shown in Fig.~\ref{fig:numerical}\ plotting $\ln V$, $\ln C$ versus
distance.  Apart from a saturation due primarily to round-off error,
both correlators clearly display exponential decay, $\ln C_{ij},\ln
V_{ij} \sim -|r_i-r_j|/\xi$ with apparently the same correlation
length $\xi \approx 1$.

Short-range exponentially decaying correlations in $C_{ij}$ establish
the absence of spin order, but do not preclude broken translational
symmetries such as plaquette or bond order.  The exponential decay of
the vison correlator $V$, however, implies that the phase is necessarily
fractionalized with deconfined $S_z=1/2$ spinon excitations,
regardless of the presence or absence of broken translational
symmetries.  The exponential spatial decay of $V_{ij}$ is suggestive
of a vison gap, i.e. the existence of a minimum energy required to
excite a vison.  A vison gap, however, strictly speaking requires
exponential decay of the vison correlator in {\sl imaginary time}, not 
in space at equal time.  Conceivably, the latter condition could occur 
in the absence of a vison gap, provided that these visons were
localized.  Given the peculiarities of the present model, we desire a
direct argument for a vison gap.  
Fortunately, such an argument can also be made using the
properties of the RK point.  Our construction closely follows and only 
slightly refines an argument used in Ref.~\onlinecite{Nayak}.
We consider a family of models defined by singling out a single ``central''
triangle of the Kagome lattice, and the associated three bow-tie ring
moves centered on the three sites of this triangle.  For these three
ring terms, we let $u'_4=J'_{\rm ring}$ vary independently of $u_4=J_{\rm
  ring}$ for all other bow-ties.  Independent of the choice of
$J'_{\rm ring}
\geq 0$, the original, translationally-invariant $|0\rangle$ state remains 
a zero energy ground-state wavefunction.  However, for the special
case $J'_{\rm ring}=0$, we can find an additional exact zero-energy eigenstate,
\begin{equation}
  |v\rangle = \hat{v}_i |0\rangle,
\end{equation} 
where $\hat{v}_i$ is the (string) vison creation operator emerging
from any of the sites of the central triangle.  Thus for $J'_{\rm ring}=0$,
there is no vison gap.  We interpret this result to mean that by
reducing the ring couplings on the central triangle, a vison has been
{\sl bound} to this triangle, with a binding energy that exactly
equals its gap in the bulk.  To test this hypothesis, we calculate the 
first-order energy shift as $J'_{\rm ring}$ is increased from zero to
positive values using perturbation theory.  To leading order, one
finds that 
\begin{eqnarray}
  E_v & = & \langle v|{\cal H}'|v\rangle \nonumber \\
  & = & \langle 0|\hat{v}_i {\cal H}' \hat{v}_i |0\rangle \nonumber\\
  & = & 6 J'_{\rm ring} \langle 0|\hat{P}_{\rm flip}|0\rangle,
\end{eqnarray}
where ${\cal H}'$ is the sum of ring terms for the three sites of the
triangle.  To obtain the final result, we use $\hat{v}_i S^x_j
\hat{v}_i = \pm S^x_j$, where the minus (plus) sign obtains if $j$ is
(is not) on the string.  The energy of the no-vison state $E_0=0$ for
all $J'_{\rm ring}$.  Hence the gap in this approximation is proportional to
$J'_{\rm ring}$ multiplied by the probability that any given bowtie is
flippable.  The latter probability is determined directly by the
classical Monte-Carlo procedure, and we find $\langle 0|\hat{P}_{\rm
  flip}|0\rangle \approx 0.257$, hence
\begin{equation}
  E_v \approx 1.54 J'_{\rm ring}.
\end{equation}  
Note that this is only the first-order approximation to the vison
gap.  The naive extension $J'_{\rm ring} \rightarrow J_{\rm ring}$ gives a
reasonable extrapolation (this is used in Ref.~\onlinecite{Nayak}),
but there is no obvious reason to expect it to be exact.

\section{Discussion}

We have demonstrated the equivalence of the generalized Kagome antiferromagnet in the easy-axis limit
to an XY ring model, and moreover shown that, with the addition of the
$u_4$ interaction, this model is in a topologically-ordered phase at
the RK point where $u_4=J_{\rm ring}$.  At this point, the model is a
true ``short-range'' spin liquid, insofar as there is evidently no
order or broken symmetry, and all excitations  are gapped.  But more importantly 
by computing the vison 2-point function we explicitly demonstrate
that this spin-liquid phase is fractionalized, and supports
$S^z = 1/2$ spinon excitations whose gap is $O(J_z)$.
Since the 
visons also are gapped -- with a gap of $O(J_{\rm ring})=O(J_\perp^2/J_z)$ --
this spin liquid state is resilient.

In particular, we may consider a variety of perturbations away from
the special soluble model.  Following similar arguments to
those of Moessner and Sondhi\cite{sondhi}, the spin liquid state 
remains the ground
state for $A<u_4/J_{\rm ring}<1$, where $A$ is some unknown
dimensionless number.  If $A<0$, then the fractionalized phase
persists to the pure ring model, but it is possible that an
intermediate phase (or phases) intervene(s) such that $A>0$.
Quantum Monte Carlo simulations could be useful in deciding this
question.

Perhaps more novel perturbations consist in deviations from the
condition $J_1=J_2=J_3$ imposed initially.  Again, the presence of a
complete gap in the spectrum rules out the destruction of the spin
liquid by these perturbations (provided they are weak).  It is
interesting to consider the effect of small changes in $J_1$, in
particular $J_{1z}=J_z + \delta J_{1z}$, $J_{1\perp}=J_\perp+\delta
J_{1\perp}$.  Viewing these deviations as perturbations, the change in
the easy-axis coupling can be rewritten as
\begin{eqnarray}
  \delta{\cal H}_z & = & \delta J_{1z} \sum_{\langle ij\rangle} S_i^z
  S_j^z \nonumber \\
  & = & {\frac{1}{4}}\delta J_{1z}  \sum_{\langle ij\rangle} \hat{v}_i
  \hat{v}_j, \label{eq:visonKE}
\end{eqnarray}
where the sum in the first line is taken over nearest-neighbor site of
the Kagome lattice, and is equivalent in the second line to a sum over 
hexagons that are nearest-neighbors within either the red or blue
honeycomb sublattice of the dual dice lattice.  Remarkably,
the latter form,
Eq.~\ref{eq:visonKE}\ corresponds to a vison hopping or
kinetic energy term.  Because the visons are already gapped, this
clearly will not destabilize the ground state provided the kinetic
energy gain remains small relative to the vison gap, i.e. $\delta J_{1z}
\lesssim J_{\rm ring}$.  

The change in the in-plane
exchange can be written
\begin{equation}
  \delta {\cal H}_\perp = {\frac{1}{2}}\delta J_{1\perp} \sum_{\langle
    ij\rangle} \left( S_i^+ S_j^- + S_i^- S_j^+\right). \label{eq:spinonKE}
\end{equation}
For any given bond on the Kagome lattice, the associated term in
Eq.~\ref{eq:spinonKE}\ raises $S_{\mbox{\small\hexagon}}^z$ of one hexagon 
by $+1$ and lowers $S_{\mbox{\small\hexagon}}^z$ of another by $-1$.
Clearly, acting upon the ground state, this takes the system outside
the low-energy manifold of $S_{\mbox{\small\hexagon}}^z=0$.  Hence due to
the large spin gap, it generates only weak second-order virtual
processes that renormalize couplings of the effective ring model.
However, its effects are more interesting on some of the excited
states.  In particular, for a single spinon excitation, one has for
$\delta J_{1\perp}=0$ a single magnetized hexagon with
$S_{\mbox{\small\hexagon}}^z=\pm 1$.  The spinon is completely localized,
and there is an associated degeneracy of these excited states
reflecting translational invariance due to the arbitrariness of which
hexagon is magnetized.  For $J_{1\perp}\neq 0$, this degeneracy is
lifted, since Eq.~\ref{eq:spinonKE}\ allows the magnetized hexagon to
move.  Thus Eq.~\ref{eq:spinonKE}\ has the effect of giving the
spinons some kinetic energy, and the associated states broaden into a
band.  Indeed, it is possible to formally rewrite
Eq.~\ref{eq:spinonKE}\ explicitly as a spinon hopping term.  As above, 
because of the existing spinon gap, the ground state is expected to be 
stable to this perturbation for $\delta J_{1\perp} \lesssim J_z$.

We conclude with a comparison of our results to some related
discussions in the literature.  An interesting aspect of the spinons
in the generalized Kagome antiferromagnet we have considered is that they are {\sl bosonic}.  Despite 
the close relation of the topologically ordered state described here
to a Gutzwiller-projected superfluid, this is 
in contrast to what is obtained by such projections on SU(2)-invariant 
superconducting states\cite{Gutzwiller}, as are naturally suggested by work arising
from various slave-fermion theories\cite{slave}.  As in our work,
the large $N$ approaches to the spin liquid\cite{LargeN} also find bosonic spinons.  

One of the most intriguing aspects of the numerical results on the
spin-$1/2$ nearest-neighbor Kagome antiferromagnet\cite{kagnum} is the
proliferation of a large number of very low-energy singlet
excitations.  Our approach does not shed too much light on this
phenomenon, since the spin liquid ground state found here is in fact
fully gapped.  It is, however, true that in the easy-axis limit
considered above
the visons (which are ``singlets''
under U(1) rotations about the $S^z$ axis) have a much smaller gap
($J_{\rm ring} =J_\perp^2/J_z \ll J_z$) than the spinons.  Moreover,
it is natural to expect in our effective ring model that as the ratio
$u_4/J_{\rm ring}$ is decreased, some confinement transition should
occur.  At such a confinement transition, provided it is second order, 
the vison gap must vanish.  Should the pure ring model lie near to
this critical point, one would indeed expect a large number of
low-lying singlet
excitations, which on the deconfined side of the critical point are
understood as weakly gapped visons.

In light of these results it should be interesting to look at other
bosonic ring models, well away from the integrable RK point.  For
instance, the properties of the pure XY ring model,
Eq.~\ref{eq:Sring}, defined on four-site plaquettes on diverse
lattices (Kagome, square, triangular...) are very poorly understood.
With the insight that such terms strongly favor vortex pairing, these
seem excellent candidate models that might exhibit quantum number
fractionalization.  A variety of numerical\cite{Doug}\ and novel analytical
techniques\cite{unpub}\ might profitably be applied to these systems.

We are grateful to T. Senthil and Doug Scalapino for scintillating
discussions.  L.B. was supported by NSF grant DMR-9985255,
  and the Sloan and Packard foundations,  M.P.A.F. by
  NSF Grants DMR-97-04005, DMR95-28578 and PHY94-07194, and S.M.G. by
  NSF Grant DMR-008713.



\begin{thebibliography}{99}

\bibitem{PWA} P.W. Anderson, Science, {\bf 235}, 1196 (1987).

\bibitem{Sachdevbook} See for example, ``Quantum Phase Transitions", by Subir Sachdev, (Cambridge, 1999).

\bibitem{RKdimer} D.S. Rokhsar and S. Kivelson,  Phys. Rev. Lett. {\bf 61}, 2376 (1988).

\bibitem{RVB} G. Baskaran and P.W. Anderson, Phys. Rev. B{\bf 37}, 580 (1988); S. Kivelson, D.S. Rokhsar, and J. Sethna, Phys. Rev. {\bf B35}, 8865 (1987). 

\bibitem{slave} I. Affleck and J.B. Marston, Phys. Rev. {\bf B37}, 3774 (1988);
L. Ioffe and A. Larkin, Phys. Rev. {\bf B 39}, 8988 (1989); 
P.A. Lee and N. Nagaosa, Phys. Rev. {\bf B45}, 966 (1992); 
P.A. Lee, Physics C {\bf 317-318}, 194 (1999) and references therein.

\bibitem{LargeN} N. Read and S. Sachdev, Phys. Rev. Lett. {\bf 66}, 1773 (1991);
S. Sachdev and N. Read, Int. J. Mod. Phys. {\bf B5}, 219 (1991). 

\bibitem{oldtop} N. Read and B. Chakraborty, Phys. Rev. {\bf B40}, 7133 (1989);
S. Kivelson, Phys. Rev. {\bf B39}, 259 (1989); X.G. Wen, Phys. Rev. {\bf B44}, 2664 (1991); C. Mudry and E. Fradkin,
Phys. Rev. {B49}, 5200 (1994). 
 

\bibitem{newtop} T. Senthil and M.P.A. Fisher, Phys. Rev. B{\bf 63},
  134521 (2001); Phys. Rev. Lett. {\bf 86}, 292 (2001)and references therein.

\bibitem{Wentop} X.G. Wen and Q.Niu, Phys. Rev. {\bf B41}, 9377 (1990). 

\bibitem{Z2} T. Senthil and M.P.A. Fisher, Phys. Rev. B{\bf 62}, 7850 (2000);
Phys. Rev. Lett. {\bf 86}, 292 (2001).

\bibitem{vortpair} L. Balents, M.P.A. Fisher, and C. Nayak, 
Phys. Rev. {\bf B60 },1654 (1999); 
Phys. Rev. B{\bf 61}, 6307 (2000).

\bibitem{kagnum} P. Lecheminant, B. Bernu, C. Lhuillier, L. Pierre and
  P.  Sindzingre, Phys. Rev. B{\bf 56}, 2521 (1997); C. Waldtmann, H.
  Everts, B. Bernu, P. Sindzingre, C. Lhuillier, P. Lecheminant and L.
  Pierre, Euro,. Phys. J. B{\bf 2}, 501 (1998) and references therein.

\bibitem{ringnum} G. Misguich, B. Bernu and C. Waldtmann,
Phys. Rev. B{\bf 60}, 1064 (1999); W. LiMing, P. Sindzingre and G. Misguich,
Phys. Rev. B{\bf 62}, 6372 (2000).

\bibitem{sondhi} R. Moessner and S. Sondhi, Phys. Rev. Lett. {\bf86}, 1881 (2001).

\bibitem{Gutzwiller} C. Gross, Phys. Rev. B{\bf 38}, 931 (1988);
D. Poilblanc, Phys. Rev. B{\bf 39}, 140 (1989).

\bibitem{Chalker} S.E. Palmer and J.T. Chalker, cond-mat0102447.

\bibitem{MPAFvortcond} M.P.A. Fisher, Phys. Rev. Lett. {\bf 65}, 923 (1990).

\bibitem{Kogut} J.B. Kogut, Rev. Mod. Phys. {\bf 51}, 659 (1979).

\bibitem{Nayak}  C. Nayak and K. Shtengel, Phys. Rev. B{\bf
    6406}, 4422 (2001).

\bibitem{Doug} A. Sandvik and D. Scalapino, unpublished.

\bibitem{unpub} L. Balents and M.P.A. Fisher, unpublished.


\end{thebibliography}
\end{document}